\documentclass[pre,preprint]{revtex4}
\usepackage{amsmath}
\usepackage{dcolumn}
\usepackage[xdvi]{graphicx}%
\makeatletter

\def\btt#1{\texttt{\@backslashchar#1}}%
\DeclareRobustCommand\bblash{\btt{\@backslashchar}}%
\makeatother

\newcommand{\bv}[1]{{\boldsymbol #1}}

\begin{document}

\title{Long-time tails in freely cooling granular gases}
\author{Hisao Hayakawa\footnote{e-mail: hisao@yukawa.kyoto-u.ac.jp}
and Michio Otsuki\footnote{e-mail: otsuki@yukawa.kyoto-u.ac.jp} }
\address{Yukawa Institute for Theoretical Physics, Kyoto University,  Kitashirakawaoiwake cho, Sakyo,
Kyoto 606-8502, JAPAN}
\begin{abstract}

The long-time behavior of the current 
auto-correlation functions for the velocity, the shear stress and 
the heat flux is investigated in freely cooling granular gases. 
It is found that the correlation functions for the velocity and 
the shear stress have the long-time tails obeying $\tau^{-d/2}$, 
while the correlation function for heat flux decays as 
  $\tau^{-(d+2)/2} \exp(-\zeta^* \tau)$ with
the dimensionless cooling rate $\zeta^*$, the spatial dimension $d$ 
and the scaled time $\tau$ in terms of the collision frequency.
The result of our numerical simulation of the freely cooling granular gases
is consistent with the theoretical prediction.
\end{abstract}
\pacs{45.70.-n,83.10.Pp,05.40.-a}
\maketitle

\section{Introduction}

One of the central concerns in granular physics is to know the rheological properties  of granular fluids \cite{aronson}. 
It is believed that 
rapid granular flows for relatively dilute granular gases can be
described by a set of hydrodynamic equations 
derived from the kinetic theory \cite{poeschel,brilliantov04,goldhirsch}.  
Most of theories assume the molecular chaos ansatz which can be only justified for dilute gases. 
For granular gases with finite density, inelastic Enskog equation has been used
\cite{JR,JR3d85,Garzo98,lutsko05} 
and the theoretical predictions well recover 
the results of simulations and experiments \cite{mitarai05,Xu03,saitoh}.
It is, however, well known that correlated collisions cause significant changes in the form of
constitutive relations for molecular fluids. 
It is important to include effects of correlated collisions for the
description of granular fluids, though so far there are few such studies.

In a recent paper, Saitoh and Hayakawa \cite{saitoh} demonstrate the relevancy of 
the kinetic approach to describe sheared granular fluids \cite{JR}. In spite of their success, there are  some unclear points. For example, they rely on the kinetic theory for 
uniform cooling granular gases, but this approach may not be valid for the description of shear flows.  In fact,  we have recognized the significant differences between 
freely cooling granular gases and sheared granular flows: (i) there is a long-range velocity correlation in freely cooling states \cite{van Noije},
but we do not find any evidence of the equal-time long-range correlation in sheared granular flows \cite{namiko07}, 
(ii) the basic solution of the inelastic Boltzmann equation in sheared granular flows \cite{goldhirsch98,lutsko04} is completely different from that of homogeneous
cooling states \cite{vannoije98}.

In general, the long-time tails obeying $t^{-d/2}$ with the time $t$
and the spatial dimension $d$ in the auto-correlation functions play
important roles for elastic gases
\cite{alder,ernst71,Pomeau}.
In fact, it is known that  the transport coefficients in two-dimensional systems diverge in the thermodynamic limit \cite{Pomeau71,Murakami03}, 
and they have the logarithmic 
singularity in the virial expansions even in three dimensional cases \cite{kawasaki-oppenheim} because of the long-time tails.
Thus, if there are the long-time tails in granular fluids, 
we may need to change the transport coefficients determined 
by the inelastic Boltzmann equation \cite{brilliantov04} and Enskog equation 
\cite{JR,JR3d85,Garzo98,lutsko05}.
In a recent paper, Kumaran \cite{kumaran} indicates the suppresses of the long-time tail in sheared granular flows as $t^{-3d/2}$,
while Ahmad and Puri \cite{ahmad} suggest the existence of the long-time tail 
as $\tau^{-1}$ with the scaled time $\tau$ by the collision frequency
in freely cooling granular gases from their two-dimensional simulation. Therefore, we need to clarify
what is the true story of the long-time tails in granular fluids and whether the logarithmic divergences of the transport coefficients  exist in two-dimensional granular gases. 

 In this paper, we discuss whether there are the long-time tails 
 in freely cooling granular fluids.
First, we demonstrate that the conventional long-time tails for the velocity auto-correlation function 
and the shear stress auto-correlation function 
exist, while  we predict that the auto-correlation function for the heat flux decays exponentially from our consideration of the
hydrodynamic fluctuations around the uniform cooling state. 
Second, we show the results of our extensive numerical simulations for 
two-dimensional freely cooling granular 
gases, which are almost consistent with our theory. Finally, we will discuss the non-trivial relations between the diffusion coefficients and the auto-correlation functions
in granular fluids.

\section{theoretical analysis} 
\label{ana:sec}

In this section, we calculate the behavior
of the time correlation functions in the freely cooling system based on the method developed in ref. \cite{ernst71}.
This section consists of four subsections associated with five appendices.

\subsection{Model}


The system considered in this paper consists of $N$ identical smooth and hard spherical particles with the mass
$m$ and 
the diameter $\sigma$ in the volume $V$. 
The position and the velocity of the $i$-th particle at time $t$
are $\bv{r}_i(t)$ and $\bv{v}_i(t)$, respectively.  
The particles collide instantaneously with each other
with a coefficient of restitution $e$ less than unity, where we do not consider 
the effects of oblique impacts and the collision speed dependence 
in $e$ \cite{walton,labous97,kuninaka03,kuwabara,brilliantov96,morgado97}.
When the particle $i$ with the velocity $\bv{v}_i$ collides with the particle  $j$ with $\bv{v}_j$,
the post-collisional velocities $\bv{v'}_i$ and $\bv{v'}_j$ are respectively given by
\begin{eqnarray}\label{eq1}
\bv{v'}_i & = & \bv{v}_i  - \frac{1}{2} (1+e) (\bv{\epsilon} \cdot
\bv{v}_{ij}) \bv{\epsilon}, \\
\bv{v'}_j & = & \bv{v}_j  + \frac{1}{2} (1+e) (\bv{\epsilon} \cdot
\bv{v}_{ij})\bv{\epsilon},
\end{eqnarray}
where $\bv{\epsilon}$ is the unit vector parallel to the relative position of the
two colliding particles at contact, and $\bv{v}_{ij}=\bv{v}_{i}-\bv{v}_{j}$.

\subsection{Basic formulae}

Here, we introduce the correlation functions as
\begin{eqnarray}
C_D(t_0,t) &=& \lim_{V\to\infty}\frac{1}{V} \sum_{i=1}^N \langle  
v_{ix}(t_0) v_{ix}(t+t_0) \rangle_{t_0}, \\
C_\eta(t_0,t) &=& \lim_{V\to\infty}\frac{1}{V}\langle J_{\eta}(t_0)
J_{\eta}(t+t_0) \rangle_{t_0}, \label{eq4} \\
C_\lambda(t_0,t) &=& \lim_{V\to\infty}\frac{1}{V}\langle 
J_{\lambda}(t_0) J_{\lambda}(t+t_0) \rangle_{t_0}, \label{eq5}
\end{eqnarray}
where $J_\eta(t)$ and $J_\lambda(t)$ are  the shear stress
and the heat flux at time $t$, respectively. 
Here,  $t_0$ is the time when we start the measurement
and $\left < \cdots \right > _{t_0}$ represents the ensemble average of
possible configuration at time $t_0$. 
In general,  the currents $J_{\eta}(t)$ and $J_{\lambda}(t)$ consist of the kinetic part and the potential part, respectively.
In this paper, we only consider the contribution from the kinetic part of the currents. This treatment is correct
in dilute gases. For higher density cases, we need a more sophisticated method to include the contribution from
the potential part, but the corrections only appear in the prefactor of coefficients at least for elastic gases \cite{dorfman75,ernst76}.
To be consistent with the assumption that the kinetic part is dominant, our calculation is based on the hydrodynamic
equations for dilute granular gases.
Thus, the currents in Eqs. (\ref{eq4}) and (\ref{eq5}) are respectively approximated  by their kinetic parts $J_{\eta}^K(t)$ and $J_{\lambda}^K(t)$ as
\begin{eqnarray}
J_{\eta}(t) & \simeq & J_{\eta}^K(t)\equiv \sum_i m v_{ix}(t)v_{iy}(t), \\
J_{\lambda}(t) & \simeq & J_{\lambda}^K(t)\equiv  \sum_i \frac{1}{2} 
\left ( m v_{i}^2(t) -(d+2)  T(t) \right ) v_{ix}, \label{eq7}
\end{eqnarray}
where $T(t)$ is the temperature at time $t$.
In order to simplify the calculation, we express these correlation functions
as
\begin{eqnarray}
C_\alpha(t_0,t) &=& \lim_{V\to\infty}\frac{1}{V} \langle  
N j_{\alpha}(\bv{v_1}(t_0))  J_\alpha(t+t_0) \rangle_{t_0}, \label{Cor:eq}
\end{eqnarray}
where $\alpha = D, \eta, \lambda$ and
\begin{eqnarray}
j_{D}(\bv{v_1}) & = & v_{1x}, \\
j_{\eta}(\bv{v_1}) & = & m v_{1x}v_{1y}, \\
j_{\lambda}(\bv{v_1}) & = & \frac{1}{2} (m v_1^2 - (d+2)  T) v_{1x}, \\
J_{D} & = & j_{D}(\bv{v_1}). 
\end{eqnarray}

In the following argument, the calculation of the correlation functions
will be separated into two steps. At the first step, we decompose the average in Eq. (\ref{Cor:eq})
into the partial averages over spatially nonuniform systems 
 where the particle 1 has a given velocity $\bv{v}_1(t_0)=\bv{v}_0$ and its initial position 
is constrained to the neighborhood $\bv{r}_0$ by the smeared out probability density $W(\bv{r}_1(t_0)-\bv{r}_0)$. 
Thus, the decomposed initial weight function is given by
$\delta(\bv{v}_0 - \bv{v}_{1}(t_0)) W(\bv{r}_0 - \bv{r}_{1}(t_0))$, where
$W(\bv{r})$ satisfies the normalization
\begin{eqnarray}
\int d\bv{r} W(\bv{r}) = 1. \label{W:def}
\end{eqnarray}
At the second step, we further average quantities over $\bv{v}_0$ and $\bv{r}_0$ in the given ensemble of granular fluids.

Following this procedure, we rewrite Eq. (\ref{Cor:eq}) as
\begin{eqnarray}
C_\alpha(t_0,t) &=& \lim_{V\to\infty}\frac{1}{V} 
\int d\bv{r}_0 \int d\bv{v}_0
j_{\alpha}(\bv{v}_0) \nonumber \\ 
& & \times \int d \bv{r} 
 \langle  N \bar{J}_\alpha(\bv{r},t+t_0) 
 \delta(\bv{v}_0 - \bv{v}_{1}(t_0))
 W(\bv{r}_0 - \bv{r}_{1}(t_0))\rangle_{t_0}, 
 \label{C1:eq}
\end{eqnarray}
where we introduce the microscopic current densities as
\begin{eqnarray}
\bar{J}_{D}(\bv{r},t+t_0) & = & j_{D}(\bv{v_1}(t+t_0)) \delta(\bv{r} - \bv{r}_{1}(t+t_0)), 
\nonumber \\
\bar{J}_{\eta}(\bv{r},t+t_0) & = & \sum_i m v_{ix}(t+t_0)v_{iy}(t+t_0) \delta(\bv{r} - \bv{r}_{i}(t+t_0)), \nonumber \\
\bar{J}_{\lambda}(\bv{r},t+t_0) & = & \sum_i \frac{1}{2} 
\left ( m v_{i}^2(t+t_0) -(d+2)  T(t+t_0) \right ) v_{ix} 
\delta(\bv{r} - \bv{r}_{i}(t+t_0)). \label{J}
\end{eqnarray}
Here, we define the conditional average $\left < F(\bv{r},t+t_0) \right >_{t_0,c}$ 
of an arbitrary function $F(\bv{r},t+t_0)$ as
\begin{eqnarray}
\left < N F(\bv{r},t+t_0)  
\delta(\bv{v}_0 - \bv{v}_{1}(t_0)) W(\bv{r}_0 - \bv{r}_{1}(t_0))\right >_{t_0}  
& \equiv &
\left < 
\delta(\bv{v}_0 - \bv{v}_{1}(t_0)) W(\bv{r}_0 - \bv{r}_{1}(t_0))\right >_{t_0} 
\left < F(\bv{r},t+t_0)  \right >_{t_0,c} \nonumber \\
& = &
f_0(t_0,v_0) \left < F(\bv{r},t+t_0) \right >_{t_0,c},
\end{eqnarray}
where $f_0(t_0,v_0)$ is the one-particle velocity distribution function at the initial time $t_0$.
The choice of $f_0(t_0,v_0)$ is not trivial. If we are interested in the relaxation process starting from 
an equilibrium state, $f_0(t_0,v_0)$ can be the Maxwell-Boltzmann distribution. This choice, however, cannot
be used for the case starting from the middle of a cooling state. If we assume that
the base state is in a homogeneous cooling state (HCS), it is natural to adopt $f_0(t_0,v_0)$ is the approximate solution
of HCS as
\begin{eqnarray}
f_0(t_0,v_0) = n_H \left[ \frac{ m}{2\pi  T_0(t_0)} \right] ^{d/2}
e ^{ -c^2}  \{ 1+a^H_2S_2(c^2) 
\}, \label{f0}
\end{eqnarray}
where $T_0(t_0)$ is the initial temperature at $t_0$. 
In HCS, $a_2^H$ is approximately described as
\begin{equation}
a^H_2=\frac{16(1-e)(1-2e)}{9+24d+8de-41 e+30(1-e)e^2}
\end{equation}
determined from the first Sonine expansion with 
$c=v\sqrt{m/2T_0(t_0)}$ and
$S_2(x)=x^2/2-(d+2)x/2+d(d+2)/8$ \cite{vannoije98}. 
It should be noted that a more sophisticated method developed by
Brilliantov and P\"{o}schel \cite{BP06} which includes the higher expansion term $a^H_3S_3(c^2)$ with $S_3(x)=-x^3/6+(d+4)/4x^2-(d+2)(d+4)x/8+d(d+2)(d+4)/48$
does not give significant differences in the statistical properties of 
HCS from that by van Noije and Ernst \cite{vannoije98,ahmad}.
Hence, we neglect the higher order terms such as $a^H_3$ for later discussions.

With the aid of  this conditional average, Eq. (\ref{C1:eq}) is reduced to
\begin{eqnarray}
C_\alpha(t_0,t) &=& \lim_{V\to\infty}\frac{1}{V} \int d\bv{r}_0 \int d\bv{v}_0
j_{\alpha}(\bv{v}_0) f_0(t_0,v_0) \int d \bv{r} 
 \left <  \bar{J}_\alpha(\bv{r},t+t_0) \right >_{t_0,c}. 
 \label{C2:eq}
\end{eqnarray}
We rewrite $\left <  \bar{J}_\alpha(\bv{r},t+t_0) \right >_{t_0,c}$ furthermore.
For $\alpha=\eta, \lambda$, from Eq. (\ref{J}), 
$\langle \bar{J}_{\alpha}(\bv{r},t+t_0) \rangle_{t_0,c}$ can be approximated by
\begin{eqnarray}
\left <  \bar{J}_\alpha(\bv{r},t+t_0) \right >_{t_0,c} 
& = & 
\int d \bv{v} j_\alpha(\bv{v}) 
\left <  \sum_i \delta(\bv{r}_i(t) - \bv{r}(t)) 
\delta(\bv{v}_i(t) - \bv{v}(t))\right >_{t_0,c}  \nonumber \\
& \simeq & 
\int d \bv{v} j_\alpha(\bv{v})  f_l(\bv{r},\bv{v},t). \label{j:eq}
\end{eqnarray}
Here, we replace
$\left <  \sum_i \delta(\bv{r}_i(t+t_0) - \bv{r}) 
\delta(\bv{v}_i(t+t_0) - \bv{v})\right >_{t_0,c}$ by the local scaling distribution
function \cite{vannoije98}
\begin{eqnarray}
f_l(\bv{r},\bv{v},t+t_0) = n(\bv{r},t+t_0) \left[\frac{ m}{2\pi  T(\bv{r},t+t_0)}\right]^{d/2}
e^{-\tilde{c}^2} 
\{1+a_2S_2(\tilde{c}^2) + \sum_{p=3}^{\infty} a_p S_p (\tilde{c}^2)\}, 
\label{fl}
\end{eqnarray}
where $\tilde{c}=\sqrt{m |\bv{v} - \bv{u}(\bv{r},t+t_0)|^2 / (2 T(\bv{r},t+t_0))
}$ 
It should be noted that $a_2$ is different from $a_2^H$ and $a_p$ with $l\geq 3$
may be relevant for inhomogeneous cooling state.
The replacement of $\left <  \sum_i \delta(\bv{r}_i(t+t_0) - \bv{r}) 
\delta(\bv{v}_i(t+t_0) - \bv{v})\right >_{t_0,c}$
by $f_l$ might be crucial. However, (i) in general, the velocity distribution
function (VDF)
can be expanded around the local Maxwellian, (ii) the transport coefficients
and the velocity correlations can be determined from the lower moments
of VDF and (iii) the long-time tails do not depend on the choice of $f_l$.
From Eq. (\ref{J}), 
$\langle \bar{J}_{D}(\bv{r},t+t_0) \rangle_{t_0,c}$ can be approximated by
\begin{eqnarray}
\left <  \bar{J}_D(\bv{r},t+t_0) \right >_{t_0,c} 
& = & 
\int d \bv{v} j_\alpha(\bv{v}) 
\left <  \delta(\bv{r}_1(t) - \bv{r}(t)) 
\delta(\bv{v}_1(t) - \bv{v}(t))\right >_{t_0,c}  \nonumber \\
& \simeq & 
\int d \bv{v} j_\alpha(\bv{v})  f_s(\bv{r},\bv{v},t), \label{j_s:eq}
\end{eqnarray}
where $f_s(\bv{r},\bv{v},t)$ is the probability density of
finding the tracer particle $1$ at position $\bv{r}$
with velocity $\bv{v}$, which is the same as Eq. (\ref{fl}) with
$n(\bv{r},t)$ is replaced by the probability density of a tracer particle
$P(\bv{r},t)$ \cite{ernst71}.

From the integration over $\bv{v}_0$ in Eqs. (\ref{j:eq}),
we rewrite the current densities as
\begin{eqnarray}
\left <  \bar{J}_D(\bv{r},t+t_0) \right >_{t_0,c}  & \simeq &
u_x(\bv{r},t+t_0)  P(\bv{r},t+t_0), \label{JD}\\
\left <  \bar{J}_\eta(\bv{r},t+t_0) \right >_{t_0,c}  & \simeq &
m n(\bv{r},t+t_0) u_x(\bv{r},t+t_0)  u_y(\bv{r},t+t_0), \label{Je}\\
\left <  \bar{J}_\lambda(\bv{r},t+t_0) \right >_{t_0,c}  & \simeq &
\frac{1}{2} (d+2)   n(\bv{r},t+t_0) (T(\bv{r},t+t_0)-T) u_x(\bv{r},t+t_0)
\nonumber \\
& & + \frac{1}{2} m n(\bv{r},t+t_0)  u^2(\bv{r},t+t_0) u_x(\bv{r},t+t_0).\label{Jl}
\end{eqnarray}
Substituting these equations into Eq. (\ref{C2:eq}), 
we rewrite the correlation functions in terms of the hydrodynamic fields.
When the linearized hydrodynamics is used, it is known that cubic terms such as the second term in the right hand side of
Eq. (\ref{Jl}) are negligible \cite{ernst71}.

\subsection{Hydrodynamic equations}

The time evolution of the hydrodynamic fields is described by the 
following equations \cite{Brey},
\begin{eqnarray}
\partial_t n + \bv{\nabla} \cdot (n\bv{u}) & = & 0, \label{neq}\\
\partial_t \bv{u} + \bv{u} \cdot \bv{\nabla} \bv{u} 
+ (nm)^{-1} \bv{\nabla} \cdot \bv{\Pi}& = & 0,  \label{ueq}\\
\partial_t T + \bv{u} \cdot \bv{\nabla} T 
+ 2 (dn)^{-1} (\bv{\Pi}:\bv{\nabla}\bv{u} 
  -\lambda \nabla^2  T - \mu \nabla^2 n)
+T \zeta & = & 0, \label{Teq} \\
(\partial_t  - D \nabla^2)  P(\bv{r},t) & = & 0, \label{Peq}
\end{eqnarray}
where $\zeta$ proportional to $1-e^2$ is the cooling rate due to inelastic collisions, and $D$ is the diffusion constant.
We should note that the diffusion equation is introduced to describe the diffusion process of the tracer particle $1$.
 
The pressure tensor $\bv{\Pi}$ for dilute gases is given by
\begin{eqnarray}
\Pi_{ij} & = & nT \delta_{ij} - \eta \left (\nabla_i u_j + \nabla_j u_i - 
\frac{2}{d} \delta_{ij} \nabla_k u_k \right ).
\end{eqnarray}
Here, the viscosity $\eta$, the heat conductivity $\lambda$, and the transport coefficient associated with the density gradient $\mu$,
the cooling rate $\zeta$ and the diffusion coefficient $D$ can be non-dimensionalized as
\begin{eqnarray}
\eta & = & \eta_0 \eta^*, \\
\lambda & = & \lambda_0 \lambda^*, \\
\mu & = & \frac{T\lambda_0}{n} \mu^*, \\
\zeta & = & \frac{nT}{\eta_0} \zeta^*, \\
D & = & \frac{2 \eta_0 }{m n} D^*,
\end{eqnarray}
where
\begin{eqnarray}
\eta_0 & = & \frac{2+d}{8} \Gamma(d/2) \pi ^{-\frac{d-1}{2}}(mT)^{1/2}
\sigma^{-(d-1)}, \\
\lambda_0 & = & \frac{d(d+2)^2}{16(d-1)} \Gamma(d/2) \pi ^{-\frac{d-1}{2}}
(\frac{T}{m})^{1/2} \sigma^{-(d-1)},
\end{eqnarray}
are the viscosity and the heat conductivity in the dilute elastic gas \cite{Brey}, 
and $\eta^*$, $\lambda^*$, $\mu^*$, $\zeta^*$, and $D^*$ are the constants 
which depend only on $e$ in dilute cases.

The hydrodynamic equations (\ref{neq}), (\ref{ueq}), (\ref{Teq}), and
(\ref{Peq}) have the set of homogeneous solutions as
\begin{eqnarray}
n(\bv{r}, t) & = & n_H, \label{ndev}\\
\bv{u}(\bv{r}, t) & = & 0, \label{udev}\\
T(\bv{r}, t) & = & T_H(t), \label{Tdev}
\end{eqnarray}
where $T_H$ obeys
\begin{eqnarray}
\frac{d T_H(t)}{dt} =  -\nu_H(t)\zeta^* T_H(t)
\end{eqnarray}
with the characteristic frequency $\nu_H=n_H T_H/\eta_0(T_H)$.
Then, we introduce the dimensionless deviations of the hydrodynamic variables as
\begin{eqnarray}
n(\bv{r}, t) & = & n_H(1+\rho(\bv{r},t)), \label{n}\\
\bv{u}(\bv{r}, t) & = & u_H(t) \bv{w}(\bv{r},t), \label{u}\\
T(\bv{r}, t) & = & T_H(t)(1+\theta(\bv{r},t)),\label{T}
\end{eqnarray}
where the variables with the subscript $H$ imply hydrodynamic variables in HCS. 
It is convenient to introduce the thermal velocity explicitly as 
\begin{eqnarray}
u_H(t)=(T_H(t)/m)^{1/2}. \label{vH}
\end{eqnarray}
Substituting Eqs. (\ref{n}), (\ref{u}), and (\ref{T}) into
Eqs. (\ref{neq}), (\ref{ueq}), and (\ref{Teq}), we obtain a set of hydrodynamic 
equations for the deviation fields $\rho, \bv{w}$, and $\theta$. To avoid to use 
time dependent coefficients,  we non-dimensionalize the set of equations 
by using
\begin{eqnarray}
\bv{\xi} & = & l_H^{-1} \bv{r}, \\
\tau & = & \frac{1}{2} \int_{t_0}^{t+t_0} ds \nu_H(s),
\end{eqnarray}
where $l_H =2u_H(t)/ \nu_H(t)$ is the characteristic length in HCS.
Here, the dimensionless time is proportional to the number of 
collisions per particle.

Therefore,  the uniform temperature decreases with the dimensionless time as 
\begin{eqnarray}
T_H(t+t_0) = T_0(t_0) \exp(-2\zeta^* \tau), \label{TH}
\end{eqnarray}
which is nothing but Haff's law \cite{Haff}.
We  linearize the set of hydrodynamic equations as
\begin{eqnarray}
\partial_\tau 
\left( 
\begin{array}{c}
 \rho_k \\
 w_{k \parallel}\\
 \theta_k \\
\end{array} 
\right)
 & = &
\left( 
\begin{array}{ccc}
0 & -ik & 0\\
-ik & \zeta^* - \frac{d-1}{d} \eta^* k^2 & -ik \\
 -  2\zeta^* - \frac{d+2}{2(d-1)} \mu^* k^2 &
 - \frac{2}{d} ik  
 & -  \zeta^* - \frac{d+2}{2(d-1)} \lambda^* k^2 \\
\end{array} 
\right)
\left( 
\begin{array}{c}
 \rho_k \\
 w_{k \parallel}\\
 \theta_k \\
\end{array} 
\right), \label{Linear:eq}\\ 
\partial_\tau \bv{w}_{k \perp}  & = & 
\left (  \zeta^* - \frac{1}{2} \eta^* k^2 \right )
\bv{w}_{k \perp}, \label{perp:eq}\\
\partial_\tau P_k  & = &  - k^2 D^* P_{k} \label{P:eq}
\end{eqnarray}
in Fourier space, where the Fourier transformed of the hydrodynamic fields
are defined by
\begin{eqnarray}
f_k(\tau) = \int d \bv{\xi} \exp(- i \bv{k} \cdot \bv{\xi}) f(\bv{\xi},\tau).
\end{eqnarray}
Here $\bv{w}_{k \perp}$ and  $w_{k \parallel}$ are the transversal velocity and 
the longitudinal component of the  velocity, which are defined as
\begin{eqnarray}
\bv{w}_{k \perp}&=&\bv{w}_{k}- (\hat{\bv{k}} \cdot \bv{w}_{k}) \hat{\bv{k}}, \\
w_{k \parallel}&=&\hat{\bv{k}} \cdot \bv{w}_{k},
\end{eqnarray}
where $\hat{\bv{k}} = \bv{k}/|\bv{k}|$.

The eigenvalues $s_\alpha$ of the matrix in Eq. (\ref{Linear:eq}) for small $k$
are calculated as
\begin{eqnarray}
s_+ & = & \zeta^* - \left ( \frac{d-1}{d}\eta^* + 
\frac{1}{d\zeta^*}\right ) k^2 + O(k^3), \nonumber \\
s_- & = & -\zeta^* - \left ( \frac{d+2}{2(d-1)}\lambda^* 
+ \frac{d+1}{d\zeta^*}\right ) k^2 + O(k^3), \nonumber \\
s_n & = &  - \frac{1}{\zeta^*} k^2 + O(k^3), \label{eigenp}
\end{eqnarray}
where $s_+, s_-$, and $s_n$ correspond to the largest, the smallest
and the neutral eigenvalues at $k=0$, respectively.
Here, we note that any sound wave proportional to $k$ 
does not exist in Eq. (\ref{eigenp}). 
This is one of the features in granular gases.
In Appendix \ref{Sound}, we discuss how the sound wave disappears
in granular gases.
In the long-time behavior,  we assume that the largest eigen-mode dominates the other modes. 
Thus, the  approximate solution of Eq. (\ref{Linear:eq})
is described as
\begin{eqnarray}
 \rho_k(t+t_0) & \simeq &
 \left ( 
 -\frac{\rho_k(t_0) k^2}{\zeta^{*2}}
 -\frac{ik w_{k \parallel}(t_0)}{\zeta^{*}}
 +\frac{(d-1)\theta_k(t_0) k^2}{d\zeta^{*2}}
 \right ) \exp(s_+ \tau), \nonumber \\
  w_{k \parallel}(t+t_0) & \simeq &
 \left ( 
 -\frac{ik \rho_k(t_0)}{\zeta^{*}}
 +w_{k \parallel}(t_0)
 +\frac{ik (d-1)\theta_k(t_0)}{d\zeta^{*}}
 \right ) \exp(s_+ \tau), \nonumber \\
 \theta_k(t+t_0) & \simeq &
 \left ( 
 \frac{(d-1)\rho_k(t_0) k^2}{d\zeta^{*2}}
 +\frac{ik (d-1) w_{k \parallel}(t_0)}{d\zeta^{*}}
 -\frac{(d-1)^2 \theta_k(t_0) k^2}{d^2\zeta^{*2}}
 \right ) 
 \exp(s_+ \tau). \label{sol:eq}
\end{eqnarray}
The details of the calculation around here are shown in Appendix 
\ref{Cal:app}.

It is easy to find the solutions of Eqs. (\ref{perp:eq}) and (\ref{P:eq}) 
\begin{eqnarray}
\bv{w}_{k \perp}(t+t_0)  & = & 
 \bv{w}_{k \perp}(t_0) e^{\left ( \zeta^* - \frac{1}{2} \eta^* k^2 \right )\tau},
 \label{wperp} \\
P_k(t+t_0)  & = &   P_{k}(t_0)e^{- k^2 D^* \tau}. \label{Pex}
\end{eqnarray}
From Eqs. (\ref{ndev}), (\ref{udev}), (\ref{Tdev}), (\ref{vH}), (\ref{TH}),
(\ref{sol:eq}) and (\ref{wperp}),
we obtain the expressions of the hydrodynamic fields as
\begin{eqnarray}
 n_k(t+t_0) & \simeq &
 \left (
 -\frac{  n_k(t_0) k^2} {\zeta^{*2}}   
 - \frac{ik n_H \sqrt{m} u_{k \parallel}(t_0)}{\zeta^* \sqrt{T_0(t_0)}}
 + \frac{(d-1)n_H T_k(t_0)k^2}{d\zeta^{*2} T_0(t_0)}
 \right )
 e^{(\zeta^{*} -bk^2) \tau},  \label{nex}\\
u_{k \parallel}(t+t_0)
 & \simeq&
 \left (
 -\frac{  i k \sqrt{T_0(t_0)} n_k(t_0)} {\zeta^{*}n_H\sqrt{m}}   
 +u_{k \parallel}(t_0)
 + \frac{i k (d-1) T_k(t_0)}{d\zeta^{*} \sqrt{m T_0(t_0)}}
 \right )
 e^{-bk^2 \tau},  \label{uparaex}\\
 T_k(t+t_0) & \simeq &
 \left (
 \frac{  (d-1) T_0(t_0) n_k(t_0) k^2} {dn_H\zeta^{*2}}   
 + \frac{ik (d-1) \sqrt{mT_0(t_0)} u_{k \parallel}(t_0)}{d\zeta^*}
 \right )e^{-(\zeta^*+bk^2)  \tau} \nonumber \\
& &  - \frac{(d-1)^2 T_k(t_0) k^2}{d^2\zeta^{*2}}
 e^{-(\zeta^*+bk^2)  \tau}, \label{Tex} \\
\bv{u}_{k \perp}(t+t_0)  & = & 
 \bv{u}_{k \perp}(t_0) e^{  - \frac{1}{2} \eta^* k^2 \tau},
\end{eqnarray}
where we introduce
$n_k = n_H \rho_k$, 
$\bv{u}_k = u_H \bv{w}_k$, 
$T_k = T_H \theta_k$, 
\begin{eqnarray}
\bv{u}_{k \perp}&=&\bv{u}_{k}- (\hat{\bv{k}} \cdot \bv{u}_{k}) \hat{\bv{k}}, \\
u_{k \parallel}&=&\hat{\bv{k}} \cdot \bv{u}_{k}, \label{uperpex}
\end{eqnarray}
and
\begin{eqnarray}\label{eq61}
b = \frac{d-1}{d}\eta^* + \frac{1}{d\zeta^*}.
\end{eqnarray}
Here, we note that $T_{k=0}$ decays exponentially as $e^{-\zeta^* \tau}$.
The origin of the exponential decay is the absence of the energy conservation law
in granular gases.

The initial values of the hydrodynamic fields depend on $\bv{v}_0$ and
$\bv{r}_0$. In this paper, following ref. \cite{ernst71}, 
we use the approximate expressions of the initial values of the 
hydrodynamic fields as
\begin{eqnarray}
n(\bv{r},t_0) & \simeq & n_H + W(\bv{r}-\bv{r_0}(t_0)), \nonumber \\
\bv{u}(\bv{r},t_0) & \simeq & \frac{\bv{v}_0}{n_H}W(\bv{r}-\bv{r_0}(t_0)), \nonumber \\
T(\bv{r},t_0) & \simeq & T_0(t_0) + \frac{m v_0^2 - dT_0(t_0)}{n_Hd}W(\bv{r}-\bv{r_0}(t_0)),
\nonumber \\
P(\bv{r},t_0) & \simeq & W(\bv{r}-\bv{r_0}(t_0)). \label{Initial:eq}
\end{eqnarray}

\subsection{Results}

Now, let us calculate the correlation functions.
Substituting Eqs. (\ref{JD}), (\ref{Je}) and (\ref{Jl}) into
Eq. (\ref{C2:eq}) with the help of Eqs. (\ref{Pex}), (\ref{uparaex}), (\ref{Tex}), and (\ref{uperpex}),
we can obtain the correlation functions.
The details of the calculation are shown in Appendices \ref{SolD:app},
\ref{Sole:app}, and \ref{Soll:app}.

First, let us calculate $C_D(t_0,\tau)$. Substituting Eq. (\ref{JD}) into
Eq. (\ref{C2:eq}),
we find 
\begin{eqnarray}
C_D(t_0,\tau) & \simeq & l_H^{-d}\int d\bv{v}_0 f_0(t_0,v_0) v_{0x} 
\int \frac{d\bv{k}}{(2\pi)^d} u_{kx}(t+t_0) P_{-k}(t+t_0) \nonumber \\
& = & 
C_D^{\perp}(t_0,\tau)+C_D^{\parallel}(t_0,\tau),
\label{CD:ex}
\end{eqnarray}
where $C_D^{\perp}(t_0,\tau)$ and $C_D^{\parallel}(t_0,\tau)$ are respectively the 
contributions from the transverse mode and the longitudinal mode for $u_{kx}(t+t_0)$. 
Their explicit definitions are respectively given by
\begin{eqnarray}
C_D^{\perp}(t_0,\tau) & = & 
l_H^{-d}\int d\bv{v}_0 f_0(t_0,v_0) v_{0x} 
\int \frac{d\bv{k}}{(2\pi)^d} u_{kx \perp }(t+t_0) P_{-k}(t+t_0),
\label{CDperp:def}\\
C_D^{\parallel}(t_0,\tau) & = & 
l_H^{-d}\int d\bv{v}_0 f_0(t_0,v_0) v_{0x} 
\int \frac{d\bv{k}}{(2\pi)^d} \hat{k}_x u_{k \parallel }(t+t_0) P_{-k}(t+t_0) .
\label{CDpara:def}
\end{eqnarray}

The result of  $C_D(t_0,\tau)$ (see Appendix \ref{SolD:app}) becomes
\begin{eqnarray}
C_D(t_0,\tau) \simeq \frac{T_0(t_0)}{dml_H^{d}}\left\{(d-1)
\left ( \frac{1}{2\pi(\eta^* + 2D^*) \tau} \right ) ^{d/2}
+ 
\left ( \frac{1}{4\pi(b + D^*) \tau} \right ) ^{d/2}\right\}, \label{CD:sol}
\end{eqnarray}
where $b$ is given in Eq. (\ref{eq61}).
The first term in Eq. (\ref{CD:sol}) is the contribution from $C_D^{\perp}(t_0,\tau)$,
 and the second term is from $C_D^{\parallel}(t_0,\tau)$. 
We should note that the second term is absent
in the elastic case \cite{ernst71}.
 The finite contribution from the second term in our problem is related to the absence of the sound wave
in granular gases. 
Thus, we obtain the asymptotic behavior of $C_D(t_0,\tau)$ 
decaying $\tau^{-d/2}$
which is consistent with the simulation by Ahmad and Puri \cite{ahmad}.

Second, 
let us calculate $C_\eta(t_0,\tau)$. Substituting Eq. (\ref{Je}) into
Eq. (\ref{C2:eq}),
we rewrite
\begin{eqnarray}\label{B6}
C_\eta(t_0,\tau) & \simeq & \frac{m^2 n_H}{ l_H^{d}}
\int d\bv{v}_0 f_0(t_0,v_0) v_{0x} v_{0y} 
\int \frac{d\bv{k}}{(2\pi)^d} u_{kx}(t+t_0) u_{-ky}(t+t_0) \nonumber \\
& = &  
C_{\eta}^{\perp\perp}(t_0,\tau)+C_{\eta}^{\parallel\perp}(t_0,\tau)+C_{\eta}^{\parallel\parallel}(t_0,\tau) ,
\label{Ce:ex}
\end{eqnarray}
where
\begin{eqnarray}
C_\eta^{\perp \perp} & = & 
m^2 n_H l_H^{-d}\int d\bv{v}_0 f_0(t_0,v_0)  v_{0x} v_{0y} 
\int \frac{d\bv{k}}{(2\pi)^d} u_{kx \perp }(t+t_0) u_{\perp -ky}(t+t_0),
\label{Cepepe:def}\\
C_\eta^{\parallel \perp} & = & 
m^2 n_H l_H^{-d}\int d\bv{v}_0 f_0(t_0,v_0) v_{0x}  v_{0y} 
\int \frac{d\bv{k}}{(2\pi)^d} 
u_{kx \perp }(t+t_0) \hat{k}_y u_{\parallel -k}(t+t_0) \nonumber \\
& & +
m^2 n_H l_H^{-d}\int d\bv{v}_0 f_0(t_0,v_0) v_{0x}  v_{0y} 
\int \frac{d\bv{k}}{(2\pi)^d} 
 \hat{k}_x u_{k \parallel }(t+t_0) u_{\perp -ky}(t+t_0),
\label{Cepepa:def} \\
C_\eta^{\parallel \parallel } & = & 
2m^2 n_H l_H^{-d}\int d\bv{v}_0 f_0(t_0,v_0)  v_{0x} v_{0y} 
\int \frac{d\bv{k}}{(2\pi)^d} 
\hat{k}_x \hat{k}_y  u_{k \parallel }(t+t_0) u_{\parallel -k}(t+t_0).
\label{Cepapa:def}
\end{eqnarray}

The result of the calculation $C_\eta(t_0,\tau)$ 
becomes (see Appendix\ref{Sole:app})
\begin{eqnarray}
C_\eta(t_0,\tau) &\simeq& \frac{2T_0(t_0)^2(1+a_2)}{d(d+2)l_H^d} 
\nonumber \\
& & \times \left ( \frac{d^2-2}{2} \left ( \frac{1}{4\pi\eta^* \tau} \right ) ^{d/2}
+d \left ( \frac{1}{2\pi(\eta^* +2b) \tau} \right ) ^{d/2}
+ \left ( \frac{1}{8\pi b \tau} \right ) ^{d/2}\right), \label{Ce:sol}
\end{eqnarray}
which also obeys $C_\eta(t_0,\tau)\sim \tau^{-d/2}$. Here, the first term comes from $C_{\eta}^{\perp\perp}(t_0,\tau)$,
the second term is from $C_{\eta}^{\parallel\perp}(t_0,\tau)$ and
the third term is from $C_\eta^{\parallel\parallel}(t_0,\tau)$.
The finite contribution from the mixing term $C_{\eta}^{\parallel\perp}(t_0,\tau)$
is also one of the characteristics in granular gases because of the absence of the sound wave.

We should note two important points. First, there should be the logarithmic corrections in $C_D(t_0,\tau)$ and $C_{\eta}(t_0,\tau)$ for $d=2$ through the self-consistent
treatment, because the coefficients include transport coefficients which show the logarithmic dependence on the system size. This is known for
the calculation of elastic gases \cite{kawasaki71}.
Second, these results suggest that the diffusion constant and the viscosity 
may depend on the system size in two-dimensional systems.
We discuss these system size dependences briefly in Sec. \ref{Summary}.

In final, let us evaluate  $C_\lambda(t_0,\tau)$. Substituting Eq. (\ref{Jl}) into
Eq. (\ref{C2:eq}),
we obtain (see Appendix\ref{Soll:app})
\begin{eqnarray}
C_\lambda(t_0,\tau) & \simeq & 
\frac{(d+2)n_H}{ 2l_H^{d}}
\int d\bv{v}_0 f_0(t_0,v_0) v_{0x}
\left ( \frac{mv_0^2}{2} - \frac{(d+2)T_0(t_0)}{2} \right ) \nonumber \\
& & \times
\int \frac{d\bv{k}}{(2\pi)^d} u_{kx}(t+t_0) T_{-k}(t+t_0) \nonumber \\
& = &  
C_{\lambda}^{\perp}(t_0,\tau)+C_{\lambda}^{\parallel}(t_0,\tau),
  \label{Cl:ex}
\end{eqnarray}
where
\begin{eqnarray}
C_\lambda^{\perp}(t_0,\tau) 
& = &  
\frac{(d+2)n_H}{ 2l_H^{d}}
\int d\bv{v}_0 f_0(t_0,v_0) v_{0x}
\left ( \frac{mv_0^2}{2} - \frac{(d+2)T_0(t_0)}{2} \right )
\nonumber \\
& & \times
\int \frac{d\bv{k}}{(2\pi)^d} 
u_{kx \perp }(t+t_0) T_{-k}(t+t_0),
  \label{Clpe:def} \\
C_\lambda^{\parallel}(t_0,\tau) 
& = &  
\frac{(d+2)n_H}{ 2l_H^{d}}
\int d\bv{v}_0 f_0(t_0,v_0) v_{0x}
\left ( \frac{mv_0^2}{2} - \frac{(d+2)T_0(t_0)}{2} \right )
\nonumber \\
& & \times
\int \frac{d\bv{k}}{(2\pi)^d} 
\hat{k}_x u_{k \parallel }(t+t_0)T_{-k}(t+t_0).
  \label{Clpa:def}
\end{eqnarray}
The calculation of $C_\lambda(t_0,\tau)$ gives 
\begin{eqnarray}
C_\lambda(t_0,\tau) 
& \simeq & -\frac{\pi (d+2)^2 (d-1)  T_0(t_0)^3 A}{2d^3m\zeta^{*2}l_H^d}
\left (  
\frac{(d- 1)e^{ - \zeta^*\tau}}{\left ( 2 \pi  (\eta^* +2b) \tau \right ) 
^{\frac{d+2}{2} }}
+ \frac{2d e^{ - \zeta^*\tau}}
 {\left ( 8 \pi  b \tau \right )^{\frac{d+2}{2} }}
 \right ), \label{Cl:sol}
\end{eqnarray}
where we introduce $A$ as
\begin{eqnarray}
A = 2(d-1) + a_2^H (9d-10). \label{A:def}
\end{eqnarray}
Here the first term in the right hand of Eq. (\ref{Cl:sol}) comes from $C_{\lambda}^{\perp}(t_0,\tau)$, and
the second term in Eq. (\ref{Cl:sol}) is from $C_{\lambda}^{\parallel}(t_0,\tau)$.
The result in Eq. (\ref{Cl:sol}) is characteristic one for granular fluids, because $C_\lambda(t_0,\tau)$ has the fast decay obeying $\tau^{-(d+2)/2}e^{ - \zeta^*\tau}$.
The absence of the long-time tail in $C_{\lambda}(t_0,\tau)$ is the result of the absence of the energy conservation law.


In our calculation, we assume  the linearity of the fluctuations around the homogeneous
cooling state. This assumption  might be  invalid to describe the long-time behavior of freely cooling granular gases. 
Indeed, it is well-known that the homogeneous cooling state is unstable,
and the system develops into inhomogeneous complicate patterns.
Hence, it is not clear whether
our analytic calculation  of the correlation functions 
is valid in the wide range of time evolution of freely cooling states.

Before we close this section, let us comment on the initial condition dependence
of our result. Through our calculation, we assume that the average is taken
over possible configuration at time $t_0$. Although we introduce $f_l$
in Eq. (\ref{fl}), we use it only to derive Eqs. (\ref{JD}), (\ref{Je}), and
(\ref{Jl}). 
Fortunately, our analysis for $C_D(t_0,t)$, 
$C_\eta(t_0,t)$, and $C_\lambda(t_0,t)$ suggests that all
correlation functions can be factorized. Namely, the correlations
are represented by the products of functions of $t_0$ and functions of $t$.
It is obvious that the important point for long time behavior is
the functions of $t$, and long-time tails are function of $t$.

\section{Numerical Simulations}

To verify the validity of our theory, in this section, we perform two-dimensional simulations to calculate
$C_D(t_0,\tau)$ and $C_\lambda(t_0,\tau)$, where we only calculate the kinetic part
of $C_\lambda(t_0,\tau)$ in Eqs. (\ref{eq5}) and (\ref{eq7}).  Here,  we do not show any result of $C_\eta(t_0,\tau)$ because
there are large fluctuations in the numerical data and the results are not
characteristic one.
To simulate the system, we employ the event driven method for hard-core particles in which the collision
rule is given by Eq. (\ref{eq1}),
where the scaled time by the collision frequency is the only
relevant time.
 The actual code of the simulations is based on 
the  efficient algorithm developed by Isobe \cite{Isobe}.
We set the volume fraction $\nu = \nu_c / 2$,
where $\nu_c$ is the close packing fraction of hard disks.
We choose the equilibrium state as the initial
state at $t=0$. In addition, we fix the time when we start the
measurement as zero.
Therefore, in this section, we replace the notation 
$C_\alpha(t_0,\tau)$ by $C_\alpha(\tau)$. 
In our simulation, $T_0(t_0)$, $m$ and $\sigma$ are set to be unity, 
and all quantities are converted to dimensionless forms.  
We should note that our numerical system is not dilute which is in contrast to the assumption of the theory.
However, as mentioned in the previous section, we expect that the contribution from the potential term is
not essentially important when the density is enough lower than the jamming transition point. Thus, we regard
the average volume fraction $\nu=\nu_c/2$ is enough "dilute".

In our simulation, the number $N$ of the particles we use
is $65536$ for the calculation of $C_D(\tau)$ with
the ensembles of $100$ different initial conditions for the case $e=1.0$
and $10$ initial conditions for the case $e<1.0$.
On the other hand, 
to suppress the large fluctuation in the calculation of $C_\lambda(\tau)$, 
we use a smaller system with $N=1024$ with the ensemble of $2250000$ initial
conditions for the case $e=1.0$ and $400000$ initial conditions for the case $e=0.95$ \cite{erpenbeck}.

\subsection{The velocity auto-correlation function}

\begin{figure}[htbp]
\begin{center}
\includegraphics[height=17em]{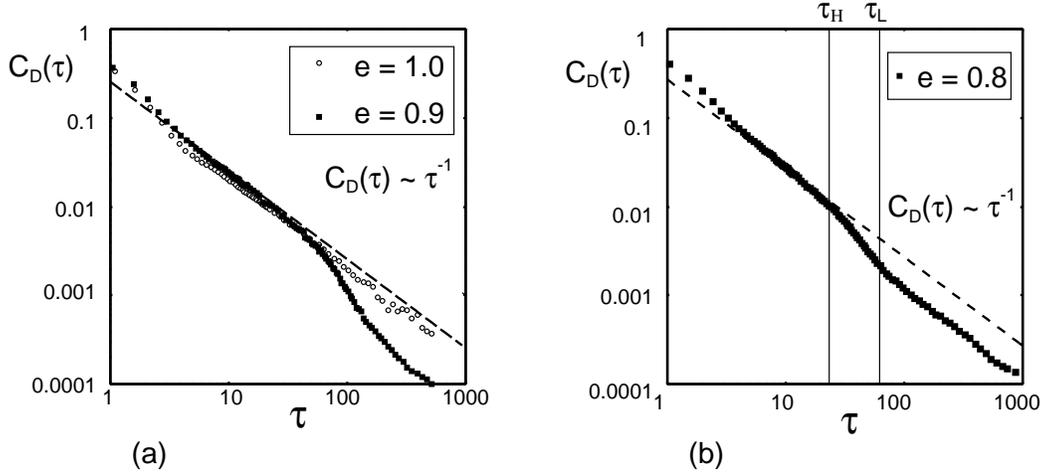}
\caption{
  (a) $C_D(\tau)$ as the function of $\tau$ with $e = 1.0, 0.9$.
  (b) $C_D(\tau)$ as the function of $\tau$ with $e = 0.8$.
  $\tau_H$ and $\tau_L$ are the time of the violation of Haff's law
  and the time starting to obey $T(\tau) \sim \tau^{-1}$, respectively.
}
\label{cr_ep1.0_0.9_0.8}
\end{center}
\end{figure}
\begin{figure}[htbp]
\begin{center}
\includegraphics[height=15em]{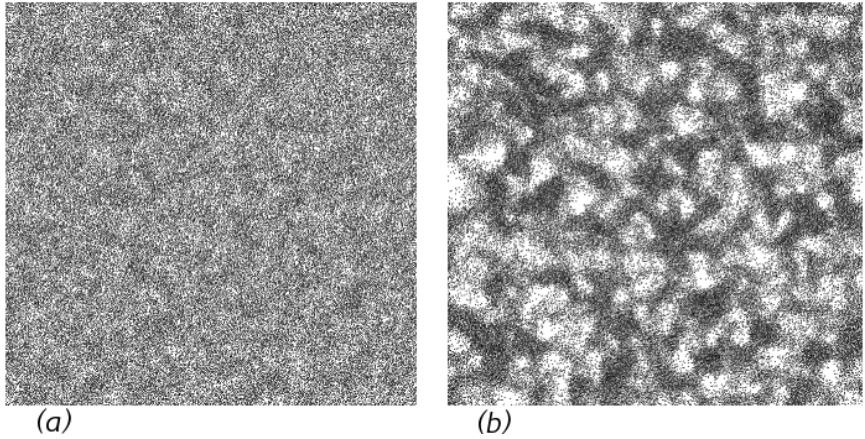}
\caption{
(a) The snapshot of the density field at $\tau=22$.
(b) The snapshot of the density field at $\tau=70$.
Both figures are obtained from simulation of $e=0.8$.
}
\label{snap}
\end{center}
\end{figure}

Figure \ref{cr_ep1.0_0.9_0.8} shows the behavior of 
$C_D(\tau)$ as the function of $\tau$ with $e = 1.0, 0.9$, and $0.8$.
For  $e=1.0$, $C_D(\tau)$ approximately obeys $\tau^{-1}$
as indicated in the previous studies \cite{Pomeau}.
For granular gases with $e<1.0$, $C_D(\tau)$ decays as $\tau^{-1}$
when the time $\tau$ is smaller than a certain threshold value $\tau_P$.
This behavior in the early regime is consistent with the result of our theory.
On the other hand, when $\tau>\tau_P$, $C_D(\tau)$ seems to decay faster than 
$\tau^{-1}$. This behavior is contrast to the prediction of our theory.
It is not surprising that our theory cannot be used  
for $\tau>\tau_P$. 
This is because the system is no longer in HCS, and
we conjecture that our theory is valid in HCS.
The fact that the system is not in HCS for $\tau>\tau_P$ is confirmed from 
Fig. \ref{snap}(a).
This figure shows the snapshot of the system with $e=0.8$ around 
$\tau=22$ which is nearly equal to $\tau_P$.
From Fig. \ref{snap}(a),
we find the existence of inhomogeneous clusters.
We also note that the threshold value $\tau_P$ decreases as $e$ decreases.

\begin{figure}[htbp]
\begin{center}
\includegraphics[height=15em]{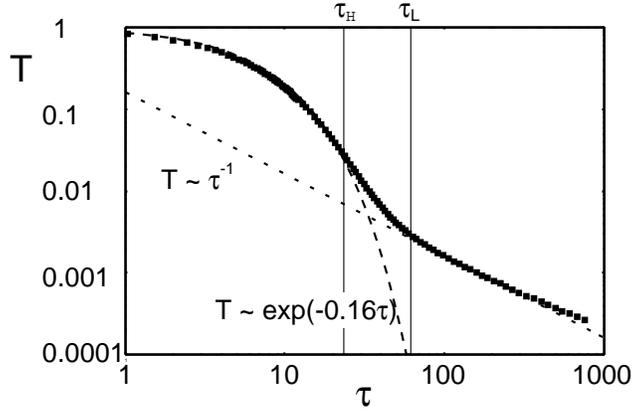}
\caption{
  $T(\tau)$ as the function of $\tau$ with $e = 0.8$.
  See the caption of Fig. \ref{cr_ep1.0_0.9_0.8} for the definitions of
  $\tau_H$ and $\tau_L$.
}
\label{T_ep0.8}
\end{center}
\end{figure}

\begin{figure}[htbp]
\begin{center}
\includegraphics[height=15em]{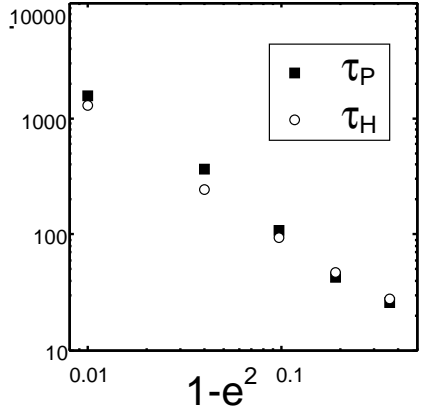}
\caption{
$\tau_{H}$ and  $\tau_{P}$
as the function of $1-e^2$.
}
\label{tau}
\end{center}
\end{figure}

We also calculate
$T(\tau)$ as the function of $\tau$ for $e=0.8$ (Fig. \ref{T_ep0.8}).  We  confirm that
the temperature decreases in terms of Haff's law (\ref{TH}) \cite{Haff} for $\tau<22$.
However, when $\tau > 22$,  Haff's law is no longer valid and the system
becomes inhomogeneous.
Here, we introduce $\tau_H$ as the time when Haff's law is violated.
Our picture that the violation of the theoretical description is related to the formation of inhomogeneous clusters
can be verified through the numerical comparison of $\tau_P$ with $\tau_H$ (Fig.\ref{tau}),
in which $\tau_H$ is close to $\tau_P$ for all $e$.
This fact  supports that
our theory is valid in HCS, but the theory may not be used when the system goes into the
inhomogeneous state.

From the data shown in Fig. \ref{cr_ep1.0_0.9_0.8},
we find another interesting behavior.
When the system goes into the late regime,
 $C_D(\tau)$ seems to recover the theoretical behavior:
$C_D(\tau)\sim \tau^{-1}$. Indeed, the simulation for $e=0.8$ clearly shows 
the existence of two regions obeying $\tau^{-1}$ as shown in  
Fig. \ref{cr_ep1.0_0.9_0.8}(b).
The time when $C_D(\tau)$ recovers the theoretical behavior
decreases as $e$ decreases.

It is interesting that the numerical result becomes consistent with the theoretical prediction  in the late regime.
We do not have any definite answer of this mysterious behavior. 
However, we note that the linearized fluctuation hydrodynamics is valid even in the late stage \cite{van Noije},
and the decay of temperature $T(t)\sim t^{-d/2}$ in the late stage can be calculated from the fluctuation of
the linearized hydrodynamics \cite{brito98}.
Here, we have confirmed that the scaled $\tau$ is almost
proportional to the real time in the late stage, which is
contrast to the relation in HCS.
Let us introduce $\tau_L$ as the time when $T(\tau)$ starts to obey 
$\tau^{-d/2}$,
as shown in Fig. \ref{T_ep0.8}.
From the comparison of $\tau_L$ and the behavior of $C_D(\tau)$ in Fig. \ref{cr_ep1.0_0.9_0.8}(b), we find
 the time that $C_D(\tau)$ starts to recover the theoretical behavior
 is nearly equal to $\tau_L$.
We also suppose that the theoretical behavior recovers
because the density becomes uniform in clusters in the late stage.
Indeed, the uniform density inside clusters may be verified in 
Fig. \ref{snap}(b) at $\tau=70$ which  is nearly equal to $\tau_L$.

\subsection{The correlation function for the heat flux}

\begin{figure}[htbp]
\begin{center}
\includegraphics[height=15em]{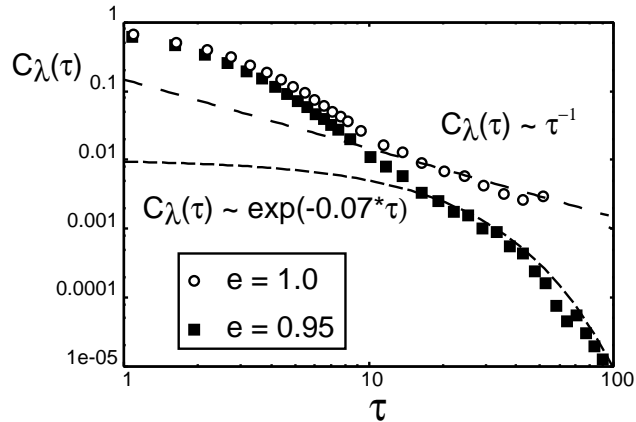}
\caption{
  $C_\lambda(\tau)$ as the function of $\tau$ with $e = 1.0, 0.95$.
}
\label{cr_heat}
\end{center}
\end{figure}

Figure \ref{cr_heat} shows the result of $C_\lambda(\tau)$ as the function of $\tau$
for  $e=1.0$ and $ 0.95$.
From this figure, we find that $C_\lambda(\tau)$ decays 
with an exponential function of $\tau$ when $e=0.95$,
while $C_\lambda(\tau)$ has a long tail obeying 
 $C_\lambda \sim \tau^{-1}$ for $e=1.0$ though the range to obey $\tau^{-1}$
 is short. 
This behavior is different from the behavior of $C_D(\tau)$ as expected from 
the theoretical prediction. 
The fluctuation of the data for $C_\lambda(\tau)$, however,  is too 
large to verify the quantitative validity of the theoretical prediction $C_\lambda(\tau) \sim - \tau^{-(d+2)/2} e^{-\zeta^* \tau}$.
It should be noted that our theory suggests negative $C_{\lambda}(\tau)$ but the simulation does not show  negative $C_{\lambda}(\tau)$.

\section{Discussion and conclusion}
\label{Summary}

In this paper, we  analytically calculate the behavior of the
time correlation functions $C_D(t_0,\tau)$, $C_\eta(t_0,\tau)$, and $C_\lambda(t_0,\tau)$.
From our calculation, we predict that
$C_D(t_0,\tau)$ and $C_\eta(t_0,\tau)$
obey $\tau^{-d/2}$, while $C_\lambda(t_0,\tau)$ decays as
$\tau^{-(d+2)/2} e^{-\zeta^* \tau}$  in the freely cooling system.
Although the behavior of $C_\lambda(t_0,\tau)$ is different from that in elastic gases, 
the behaviors of $C_D(t_0,\tau)$ and $C_\eta(t_0,\tau)$ are similar to those in the elastic cases.
We also perform the numerical simulation of the freely cooling system to verify the results of the theoretical prediction.
Through our numerical calculation of correlation functions, we find that
our prediction is valid in HCS. 
Although our prediction cannot be used in the middle regime of inhomogeneous state, 
the theoretical behavior is recovered in the late regime.

The prediction of our theory is completely different from Kumaran's prediction
\cite{kumaran}
for the sheared granular fluids in which the correlation functions
obey $t^{-3d/2}$. 
One reason of the difference between the theories comes from the 
difference of
the base states in the analyses. 
In our theory, we choose the homogeneous cooling state as
the base state, while the base state in  the sheared granular flow  is the uniform shear flow
in which the velocity gradient is uniform, the temperature and the density are uniform.

In the elastic gases, the autocorrelation functions are directly related to the transport coefficients.
Actually the diffusion
coefficient, the viscosity and the heat conductivity are respectively given by $D \propto \int_0^\infty dt C_D(t)$, 
$\eta \propto \int_0^\infty dt C_\eta(t)$, and 
$\lambda \propto \int_0^\infty dt C_\lambda(t)$ for conventional fluids.
It is known that such simple Green-Kubo formula may not be valid 
in granular gases \cite{dufty02,dufty02b,dufty02c,brey05,brilliantov05}.

The anomalous relation between the transport coefficient and the correlation
functions
in granular fluids is easily verified from the following simple argument. If we assume 
$C_D(t_0, \tau) = \left < \bv{v}(t+t_0) \cdot \bv{v}(t_0) \right >_{t_0}
= T_0(t_0) \phi(\tau)$,
then, we obtain the relation
\begin{eqnarray}\label{eq67}
D = \frac{\left < r^2(\tau) \right >}{2d \tau}
=  \frac{2T_0(t_0)}{\tau}
\int_0^{\tau} ds \phi(s) \frac{1}{2 \zeta^*} (1-e^{-2\zeta^* (\tau-s)})
\end{eqnarray}
for sufficiently large $\tau$, where $\left < r^2(\tau) \right >$
  is the mean square displacement of the tracer particle during $\tau$.
 Although the explicit expression seems to be different from the conventional Green-Kubo formula,
the logarithmic singularities in two-dimensional systems are still valid at least near $e=1$. 
In fact, the integral can be evaluated as 
$O(\tau\int_0^\tau ds\phi(s))=O(\tau \log \tau)$ for $d=2$ and 
$\tau<1/(2\zeta^*)$ with $1/(2\zeta^*)$ is proportional to $(1-e^2)^{-1}$.
For this case we might replace the cut-off time by the characteristic time depending on the linear system size $L$ such as $L^2/D$,  and $D$ may be determined
by the self-consistent way. Thus, we expect that the diffusion coefficient depends on the  system size as the result of the logarithmic singularity. 
We also indicate that Eq. (\ref{eq67}) implies that the diffusion disappears  in the limit of large $\tau$, because
the exponential term is negligible. This is also consistent with the observation in our simulation.

In this paper, we adopt Eq. (\ref{f0}) as the initial weight function which depends
on the initial temperature. 
However, this dependence may be removed 
if we map HCS onto a steady state \cite{Lutsko01}.
We expect that this mapping make our calculation clearer.

We also assume that the coefficient of
restitution $e$ is a constant in contrast to actual cases 
 \cite{walton,labous97,kuninaka03,kuwabara,brilliantov96,morgado97}.
 It may be possible to extend our results to the 
gases of viscoelastic particles \cite{Brilliantov02,Brilliantov03}.
However, we believe that
our prediction for the algebraic time dependence  
of the correlation functions
is unchanged.

We adopt, here,  the phenomenological approach developed by
Ernst et al. \cite{ernst76}. The advantages of this method
is that the calculation is simple and the physical pictures
i.e. the roles of conserved quantities are clear. This approach is successful
for the first step to indicate the relevant roles of long-time tails. However,
  the approach has the limitation in which our theoretical prediction cannot
  be used in the middle stage. To improve this, we may need a
  more fundamental approach starting from the Liouville equation \cite{Brey97}.

In conclusion, we confirm the existence of the long-time tail in the 
auto-correlation functions for the velocity and the shear stress, while it does
not exist for the heat flux. These results are consistently obtained
from the theory and simulation. Thus, the transport coefficient
based on inelastic Enskog equation should be modified for two-dimensional
systems.

\vspace*{0.5cm}

One of the authors (HH) 
thanks J. Wakou for his useful comments. 
We also appreciates the useful comments on long-time tails by M. Isobe
and H. Wada.
This study is partially supported by
Ministry of Education, Culture, Sports, Science and Technology (MEXT), 
Japan (Grant No. 18540371) and the Grant-in-Aid for the 21st century COE 
"Center for Diversity and Universality in Physics" from MEXT, Japan.
The numerical calculations were carried out on Altix3700 BX2 at YITP in Kyoto University.

\appendix

\section{The time evolution of the hydrodynamic fields}
\label{Cal:app}

In this appendix, we calculate the solution of Eq. (\ref{Linear:eq}).
The eigenvalue $s$ of the matrix in Eq. (\ref{Linear:eq}) 
satisfies the equation 
\begin{eqnarray}
& & -s \left ( s-\zeta^* + \frac{d-1}{d} \eta^*  k^2 \right )
\left ( s+\zeta^* + \frac{d+2}{2(d-1)}\lambda^*  k^2 \right ) \nonumber \\
& & - \left ( \frac{d+2}{d} s-\zeta^* + \frac{d+2}{2(d-1)}\lambda^*  k^2 
 - \frac{d+2}{2(d-1)}\mu^*  k^2 \right )k^2
 = 0. \label{eigen:eq}
\end{eqnarray}
To discuss hydrodynamic behaviors, 
we expand the eigenvalue around $k=0$ as
\begin{eqnarray}
s = s_0 + s_1 k + s_2 k^2 + \cdots. \label{s:expand}
\end{eqnarray}
Substituting this equation into Eq. (\ref{eigen:eq}),
we find that there exists three eigenvalues $s_+, s_-$, and $s_n$ as 
Eqs. (\ref{eigenp}).
Introducing the eigenvector $\bv{X}_\alpha$ corresponding 
to the eigenvalue $s_\alpha$ with the condition $|\bv{X}_\alpha|^2 = 1$, 
we find that the solution of 
Eq. (\ref{Linear:eq}) is expressed as
\begin{eqnarray}
\left( 
\begin{array}{c}
 \rho_k(t+t_0) \\
 w_{k \parallel}(t+t_0)\\
 \theta_k(t+t_0) \\
\end{array} 
\right)
 & = &
 a_+ \exp(s_+ \tau) \bv{X}_+ +  a_- \exp(s_- \tau) \bv{X}_- 
 +  a_n \exp(s_n \tau) \bv{X}_n, 
\end{eqnarray}
where we define $a_\alpha$ as
\begin{eqnarray}
a_\alpha = 
\left( \rho_k(t_0), w_{k\parallel}(t_0), \theta_k(t_0) \right)
 \cdot \bv{X}_\alpha.
\end{eqnarray}
Near $k=0$, the eigen-vector $\bv{X}_+$ can be represented by 
\begin{eqnarray}
\bv{X}_+ = 
\left (
\begin{array}{c}
 - \frac{ik}{\zeta^*} + O(k^2) \\
 1 + O(k)\\
 \frac{(d-1)ik}{d\zeta^*} + O(k^2)  \\
\end{array} 
\right),
\end{eqnarray}
and 
\begin{eqnarray}
a_+ =
-\frac{  \rho_k(t_0) k } {\zeta^{*}}
+ w_{k \parallel}(t_0)
+\frac{ (d-1) ik \theta_k(t_0)} {d\zeta^{*}} + \cdots.
\end{eqnarray}
When the time $\tau$ is enough large, the mode corresponding to the largest 
eigenvalue $s_+$ is only the relevant mode. Hence,
we obtain the approximate solution of Eq. (\ref{Linear:eq}) as Eq. (\ref{sol:eq}) in the long-time behavior.

\section{The absence of the sound wave}
\label{Sound}

It is well known that
the eigenvalue $s$ has the term proportional to $k$ which corresponds
to the sound wave for elastic gases. However, as shown in Eq. (\ref{eigenp}), 
for granular gases with $e<1.0$,
the eigenvalue $s$ does not have the term proportional to $k$.
This might be puzzling because 
the term proportional to $k$ does not appear
in the elastic limit of Eq. (\ref{eigenp}) at the first glance.

In order to understand this puzzle, let us show the 
relevant point of the calculation 
of $s$ in Appendix \ref{Cal:app}.
Substituting Eq. (\ref{s:expand}) into (\ref{eigen:eq})
with equating coefficients of each power of $k$,
we obtain the expression of $s$.

In the $0$-th and first order of $k$,
we obtain the equation
\begin{eqnarray}
s_0(s_0 - \zeta^*)(s_0 + \zeta^*)=0, \label{s0}\\
s_1^3 (3s_0^2-\zeta^{*2})=0. \label{s1}
\end{eqnarray}
From Eq. (\ref{s0}), we obtain
\begin{eqnarray}
s_0 = \alpha \zeta^*,
\end{eqnarray}
where $\alpha = 0, \pm 1$ with $\zeta^* \propto 1-e^2$.
Substituting this equation into Eq. (\ref{s1}),
we find
\begin{eqnarray}
s_1^3 (3\alpha^2 - 1)\zeta^{*2}=0.
\end{eqnarray}
For $\zeta^* \neq 0$, thus, 
$s_1$ only satisfies
\begin{eqnarray}
s_1 = 0.
\end{eqnarray}
However, for the elastic case with $\zeta^* = 0$, 
$s_1$ is not determined by this equation
but by the equation obtained from the coefficients of $k^3$
in Eq. (\ref{eigen:eq}), and we find that $s_1$ has some non-zero values.
Thus, the elastic limiti is rather singular in granular fluids.
This is the reason that the there are no sound wave when $e < 1.0$.

\section{The calculation of $C_D(t_0,\tau)$}
\label{SolD:app}
In this appendix, we calculate the correlation function Eq.
(\ref{CD:sol}).
From Eq. (\ref{CD:ex}), $C_D(t_0,\tau)$ is expressed as
\begin{eqnarray}
C_D(t_0,\tau) = C_D^{\perp}(t_0,\tau) + C_D^{\parallel}(t_0,\tau), \label{CD:ex2}
\end{eqnarray}
where $C_D^{\perp}(t_0,\tau)$ and $C_D^{\parallel}(t_0,\tau)$ are defined in Eqs. 
(\ref{CDperp:def}) and (\ref{CDpara:def}).

Substituting Eqs. (\ref{Pex}) and (\ref{uperpex}) 
into Eq. (\ref{CDperp:def}), we obtain the expression of $C_D^{\perp}(t_0,\tau)$ as
\begin{eqnarray}
C_D^{\perp} & = & 
l_H^{-d}\int d\bv{v}_0 f_0(t_0,v_0) v_{0x} 
\int \frac{d\bv{k}}{(2\pi)^d} u_{kx \perp }(t_0) P_{-k}(t_0) 
e^{-\left ( \frac{1}{2}\eta^* + D^* \right ) k^2 \tau} \nonumber \\
& = & 
\frac{1}{n_H l_H^d}\int d\bv{v}_0 f_0(t_0,v_0) v_{0x} 
\int \frac{d\bv{k}}{(2\pi)^d} (v_{0x} - \hat{k}_x \bv{v}_0 \cdot \hat{\bv{k}})
|W_k|^2
e^{-\left ( \frac{1}{2}\eta^* + D^* \right ) k^2 \tau} \nonumber \\
& \simeq & 
\frac{T_0(t_0)}{m l_H^d} 
\int \frac{d\bv{k}}{(2\pi)^d} (1-\hat{k}_x^2)
e^{-\left ( \frac{1}{2}\eta^* + D^* \right ) k^2 \tau} \nonumber \\
& = & 
\frac{(d-1)T_0(t_0)}{dm l_H^d} 
\left(
\frac{1}{2\pi(\eta^* + 2D^*)\tau}
\right )^{\frac{d}{2}}, \label{CDperp:sol}
\end{eqnarray}
where we use the relation derived from Eq (\ref{W:def}) as
\begin{eqnarray}
\lim_{k\rightarrow 0} |W_k| = 1. \label{W:app}
\end{eqnarray}
In addition, we also use  the relation 
\begin{eqnarray}
\int d\hat{\bv{k}} \hat{k}_i\hat{k}_j = \delta_{ij}/d. \label{k2rel}
\end{eqnarray}

Similarly, substituting Eqs. (\ref{Pex}) and (\ref{uparaex}) 
into Eq. (\ref{CDperp:def}) with the use of Eq. (\ref{W:app}), 
we obtain the expression for $C_D^{\parallel}(t_0,\tau)$ as
\begin{eqnarray}
C_D^{\parallel} & = & 
l_H^{-d}\int d\bv{v}_0 f_0(t_0,v_0) v_{0x} 
\int \frac{d\bv{k}}{(2\pi)^d} \hat{k}_x u_{k \parallel }(t_0) P_{-k}(t_0) 
e^{-\left ( b + D^* \right ) k^2 \tau} \nonumber \\
& = & 
\frac{1}{n_H l_H^d}\int d\bv{v}_0 f_0(t_0,v_0) v_{0x} 
\int \frac{d\bv{k}}{(2\pi)^d} 
\left ( \hat{k}_x \bv{v}_0 \cdot \hat{\bv{k}}\right )
|W_k|^2
e^{-\left ( b + D^* \right ) k^2 \tau} \nonumber \\
& \simeq & 
\frac{T_0(t_0)}{m l_H^d} 
\int \frac{d\bv{k}}{(2\pi)^d} \hat{k}_x^2
e^{-\left ( b + D^* \right ) k^2 \tau} \nonumber \\
& = & 
\frac{T_0(t_0)}{dm l_H^d} 
\left(
\frac{1}{4\pi(b + D^*)\tau}
\right )^{\frac{d}{2}}. \label{CDpara:sol}
\end{eqnarray}
From Eqs. (\ref{CD:ex2}), (\ref{CDperp:sol}) and (\ref{CDpara:sol}),
we obtain Eq. (\ref{CD:sol}).

\section{The calculation of $C_\eta(t_0,\tau)$}
\label{Sole:app}
Let us calculate $C_{\eta}(t_0,\tau)$.
From Eq. (\ref{Ce:ex}), $C_\eta(t_0,\tau)$ is expressed as
\begin{eqnarray}
C_\eta(t_0,\tau) = C_\eta^{\perp \perp} (t_0,\tau)
+ C_\eta^{\parallel \perp} (t_0,\tau)
+ C_\eta^{\parallel \parallel}(t_0,\tau), \label{Ce:ex2}
\end{eqnarray}
where $C_{\eta}^{\perp\perp}(t_0,\tau)$, $C_{\eta}^{\perp\parallel}(t_0,\tau)$ and 
$C_{\eta}^{\parallel\parallel}(t_0,\tau)$ are respectively defined by Eqs. 
(\ref{Cepepe:def}), (\ref{Cepepa:def}) and (\ref{Cepapa:def}).

Substituting Eq. (\ref{uperpex}) into Eq. (\ref{Cepepe:def}),
we obtain the expression of $C_\eta^{\perp \perp}(t_0,\tau)$ as
\begin{eqnarray}
C_\eta^{\perp \perp}(t_0,\tau) & = & 
m^2 n_H l_H^{-d}\int d\bv{v}_0 f_0(t_0,v_0)  v_{0x} v_{0y} 
\int \frac{d\bv{k}}{(2\pi)^d} u_{kx \perp }(t_0) u_{\perp -ky}(t_0)
e^{-\eta^* k^2 \tau} \nonumber \\
& = & \frac{m^2}{ n_H l_H^d}\int d\bv{v}_0 f_0(t_0,v_0)  v_{0x} v_{0y} 
\int \frac{d\bv{k}}{(2\pi)^d} 
(v_{0x} - \hat{k}_x \bv{v}_0 \cdot \hat{\bv{k}})
(v_{0y} - \hat{k}_y \bv{v}_0 \cdot \hat{\bv{k}})
|W_k|^2
e^{-\eta^* k^2 \tau} \nonumber \\
& = & \frac{T_0(t_0)^2(1+a_2^H)}{l_H^d}
\int \frac{d\bv{k}}{(2\pi)^d} 
(1+2\hat{k}_x^2\hat{k}_y^2-\hat{k}_x^2-\hat{k}_y^2)
|W_k|^2 e^{-\eta^* k^2 \tau} \nonumber \\
& = & \frac{(d^2-1)T_0(t_0)^2}{d(d+2)l_H^d}(1+a_2^H)
\left ( 
\frac{1}{4\pi \eta^* \tau}
\right )^\frac{d}{2},
\end{eqnarray}
where we use Eq. (\ref{k2rel}) and
\begin{eqnarray}
\int d\hat{\bv{k}} \hat{k}_i\hat{k}_j \hat{k}_l\hat{k}_m 
= \frac{\delta_{ij}\delta_{lm}+ \delta_{il}\delta_{jm} 
+\delta_{im}\delta_{jl}}{d(d+2)}. \label{k4rel}
\end{eqnarray}
Similarly, substituting Eqs. (\ref{uparaex}) and (\ref{uperpex}) into Eq. 
(\ref{Cepepa:def}),
we obtain the expression of $C_\eta^{\parallel\perp}(t_0,\tau)$ as
\begin{eqnarray}
C_\eta^{\parallel\perp}(t_0,\tau) & = & 
m^2 n_H l_H^{-d}\int d\bv{v}_0 f_0(t_0,v_0)  v_{0x} v_{0y} 
\int \frac{d\bv{k}}{(2\pi)^d} u_{kx \perp }(t_0) \hat{k}_y u_{\parallel -k}(t_0)
e^{-\left ( \frac{1}{2}\eta^* + b \right ) k^2 \tau} \nonumber \\
& & +
m^2 n_H l_H^{-d}\int d\bv{v}_0 f_0(t_0,v_0)  v_{0x} v_{0y} 
\int \frac{d\bv{k}}{(2\pi)^d} \hat{k}_x u_{k \parallel }(t_0) u_{k \perp y}(t_0) 
e^{-\left ( \frac{1}{2}\eta^* + b \right ) k^2 \tau} \nonumber \\
& = & \frac{m^2}{ n_H l_H^d}\int d\bv{v}_0 f_0(t_0,v_0)  v_{0x} v_{0y} 
\int \frac{d\bv{k}}{(2\pi)^d} 
(v_{0x} - \hat{k}_x (\bv{v}_0 \cdot \hat{\bv{k}}))
\hat{k}_y (\bv{v}_0 \cdot \hat{\bv{k}})
|W_k|^2
e^{-\left ( \frac{1}{2}\eta^* + b \right ) k^2 \tau} \nonumber \\
&  & + \frac{m^2 n_H}{ l_H^d}\int d\bv{v}_0 f_0(t_0,v_0)  v_{0x} v_{0y} 
\int \frac{d\bv{k}}{(2\pi)^d} 
\hat{k}_x (\bv{v}_0 \cdot \hat{\bv{k}})
(v_{0y} - \hat{k}_y (\bv{v}_0 \cdot \hat{\bv{k}}))
|W_k|^2
e^{-\left ( \frac{1}{2}\eta^* + b \right ) k^2 \tau} \nonumber \\
& = & \frac{T_0(t_0)^2(1+a_2^H)}{l_H^d}
\int \frac{d\bv{k}}{(2\pi)^d} 
(\hat{k}_x^2 +\hat{k}_y^2  - 4\hat{k}_x^2 \hat{k}_y^2)
|W_k|^2 
e^{-\left ( \frac{1}{2}\eta^* + b \right ) k^2 \tau} \nonumber \\
& = & \frac{2 T_0(t_0)^2}{(d+2)l_H^d}(1+a_2^H)
\left ( 
\frac{1}{2\pi (\eta^* +2b) \tau}
\right )^\frac{d}{2}.
\end{eqnarray}
Substituting Eq. (\ref{uparaex}) into Eq. 
(\ref{Cepapa:def}),
we obtain the expression of $C_\eta^{\parallel \parallel}(t_0,\tau)$ as
\begin{eqnarray}
C_\eta^{\parallel \parallel}(t_0,\tau) & = & 
m^2 n_H l_H^{-d}\int d\bv{v}_0 f_0(t_0,v_0)  v_{0x} v_{0y} 
\int \frac{d\bv{k}}{(2\pi)^d} \hat{k}_x \hat{k}_y 
u_{k \parallel }(t_0)u_{\parallel -k}(t_0)
e^{-2 b  k^2 \tau} \nonumber \\
& = & \frac{m^2}{ n_H l_H^d} \int d\bv{v}_0 f_0(t_0,v_0)  v_{0x} v_{0y} 
\int \frac{d\bv{k}}{(2\pi)^d} 
 \hat{k}_x\hat{k}_y (\bv{v}_0 \cdot \hat{\bv{k}})^2
|W_k|^2
e^{-2b k^2 \tau} \nonumber \\
& = & \frac{T_0(t_0)^2(1+a_2^H)}{l_H^d}
\int \frac{d\bv{k}}{(2\pi)^d} 
2\hat{k}_x^2\hat{k}_y^2 |W_k|^2 
e^{- 2b  k^2 \tau} \nonumber \\
& = & \frac{2 T_0(t_0)^2}{d(d+2)l_H^d}(1+a_2^H)
\left ( 
\frac{1}{8\pi b \tau}
\right )^\frac{d}{2}.
\end{eqnarray}
From these equations, we obtain Eq. (\ref{Ce:sol}).

\section{The calculation of $C_\lambda(t_0,\tau)$}
\label{Soll:app}
In this Appendix, we calculate $C_{\lambda}(t_0,\tau)$.
From Eq. (\ref{Cl:ex}), $C_\lambda(t_0,\tau)$ is the combination of two modes.
\begin{eqnarray}
C_\lambda(t_0,\tau) =
C_\lambda^{\perp}(t_0,\tau) 
+ C_\lambda^{\parallel}(t_0,\tau), \label{Cl:ex2}
\end{eqnarray}
where $C_{\lambda}^{\perp}(t_0,\tau)$ and $C_{\lambda}^{\parallel}(t_0,\tau)$ are respectively defined by Eqs. (\ref{Clpe:def}) and (\ref{Clpa:def}).
Since the calculation is complicated, we separate the contribution from the transverse mode and that from the longitudinal mode in the explanation.

\subsection{The contribution from the transverse mode}

Let us evaluate $C_{\lambda}^{\perp}(t_0,\tau)$ which is the contribution from the transverse mode at first. 
Substituting Eqs. (\ref{Tex}) and (\ref{uperpex}) into Eq. 
(\ref{Clpe:def}), we obtain the expression of $C_\lambda^{\perp}(t_0,\tau)$ as
\begin{eqnarray}
C_\lambda^{\perp}(t_0,\tau) 
& = &  
\frac{(d+2)n_H}{ 4l_H^{d}}
\int d\bv{v}_0 f_0(t_0,v_0) v_{0x}
\left ( mv_0^2 - (d+2)T_0(t_0) \right )
\nonumber \\
& & \times
\int \frac{d\bv{k}}{(2\pi)^d} 
u_{kx \perp }(t_0)
 e^{-\left \{ \zeta + \left ( \frac{1}{2} \eta^* +b \right )k^2 \right \} \tau}
 \nonumber \\
& & \times 
 \left (
 \frac{  (d-1) T_0(t_0) n_k(t_0) k^2} {dn_H\zeta^{*2}}   
 + \frac{ik (d-1) \sqrt{mT_0(t_0)} u_{k \parallel}(t_0)}{d\zeta^*}
 - \frac{(d-1)^2 T_k(t_0) k^2}{d^2\zeta^{*2}}
 \right ) \nonumber \\
 & = &
C_\lambda^{\perp T}(t_0,\tau) + C_\lambda^{\perp n}(t_0,\tau), \label{Clp:ex}
\end{eqnarray}
where 
we introduce
\begin{eqnarray}
C_\lambda^{\perp T}(t_0,\tau) 
& = &  -
\frac{(d+2)(d-1)^2n_H}{ 4d^2 \zeta^{*2}l_H^{d}}
\int d\bv{v}_0 f_0(t_0,v_0) v_{0x}
\left ( mv_0^2 - (d+2)T_0(t_0) \right )
\nonumber \\
& & \times
\int \frac{d\bv{k}}{(2\pi)^d} 
u_{kx \perp }(t_0)T_{-k}(t_0) k^2
 e^{-\left \{ \zeta + \left ( \frac{1}{2} \eta^* +b \right )k^2 \right \} \tau},
\label{ClpT:def} \\
C_\lambda^{\perp n}(t_0,\tau) 
& = &  
\frac{(d+2)(d-1)T_0(t_0)}{ 4d \zeta^{*2}l_H^{d}}
\int d\bv{v}_0 f_0(t_0,v_0) v_{0x}
\left ( mv_0^2 - (d+2)T_0(t_0) \right )
\nonumber \\
& & \times
\int \frac{d\bv{k}}{(2\pi)^d} 
u_{kx \perp }(t_0) n_{-k}(t_0) k^2
 e^{-\left \{ \zeta + \left ( \frac{1}{2} \eta^* +b \right )k^2 \right \} \tau}.
\label{Clpn:def}
\end{eqnarray}
Here, 
we use the fact that the term proportional to 
$u_{kx \perp }(t_0) u_{-k \parallel}(t_0)$ 
in Eq. (\ref{Tex}) does not survive in the integration over $\bv{v}_0$.

Substituting Eq. (\ref{Initial:eq}) into Eq. (\ref{ClpT:def}),
we obtain 
\begin{eqnarray}
C_\lambda^{\perp T}(t_0,\tau) 
& = &  -
\frac{(d+2)(d-1)^2n_H}{ 4d^3 n_H \zeta^{*2}l_H^{d}}
\int d\bv{v}_0 f_0(t_0,v_0) v_{0x}
\left ( mv_0^2 - (d+2)T_0(t_0) \right ) 
\nonumber \\
& & \times
\int \frac{d\bv{k}}{(2\pi)^d} 
(v_{0x} - \hat{k}_x \bv{v}_0 \cdot \hat{\bv{k}}) 
(mv_0^2 - dT_0(t_0)) |W_k|^2 k^2
 e^{-\left \{ \zeta + \left ( \frac{1}{2} \eta^* +b \right )k^2 \right \} \tau}
 \nonumber \\
& \simeq &  -
\frac{(d+2)^2(d-1)^2 T_0(t_0)^3}{ 4d^3 m\zeta^{*2}l_H^{d}}(2+a_2^H(d+10))
\int \frac{d\bv{k}}{(2\pi)^d} 
(1 - \hat{k}_x^2) 
 e^{-\left \{ \zeta + \left ( \frac{1}{2} \eta^* +b \right )k^2 \right \} \tau}
 \nonumber \\
& = &  -
\frac{\pi(d+2)^2(d-1)^2 T_0(t_0)^3}{ 2d^3 m\zeta^{*2}l_H^{d}}(2+a_2^H(d+10))
\frac{ e^{-\zeta^* \tau}}{(2\pi(\eta^* + 2b)\tau)^{\frac{d+2}{2}}}.
\label{ClpT:sol}
\end{eqnarray}
Substituting Eq. (\ref{Initial:eq}) into Eq. (\ref{Clpn:def}),
we obtain 
\begin{eqnarray}
C_\lambda^{\perp n}(t_0,\tau) 
& = & 
\frac{(d+2)(d-1)T_0(t_0)}{ 4d n_H \zeta^{*2}l_H^{d}}
\int d\bv{v}_0 f_0(t_0,v_0) v_{0x}
\left ( mv_0^2 - (d+2)T_0(t_0) \right ) 
\nonumber \\
& & \times
\int \frac{d\bv{k}}{(2\pi)^d} 
(v_{0x} - \hat{k}_x \bv{v}_0 \cdot \hat{\bv{k}}) 
|W_k|^2 k^2
 e^{-\left \{ \zeta + \left ( \frac{1}{2} \eta^* +b \right )k^2 \right \} \tau}
 \nonumber \\
& \simeq & 
\frac{(d+2)^2(d-1)T_0(t_0)^3 a_2^H}{ 4d m\zeta^{*2}l_H^{d}}
\int \frac{d\bv{k}}{(2\pi)^d} 
(1 - \hat{k}_x^2) 
 e^{-\left \{ \zeta + \left ( \frac{1}{2} \eta^* +b \right )k^2 \right \} \tau}
 \nonumber \\
& = & 
\frac{\pi(d+2)^2(d-1)^2 T_0(t_0)^3 a_2^H}{ 2d m\zeta^{*2}l_H^{d}}
\left (
\frac{ e^{-\zeta^* \tau}}{(2\pi(\eta^* + 2b)\tau)^{\frac{d+2}{2}}}
\right ).
\label{Clpn:sol}
\end{eqnarray}
Substituting Eqs. (\ref{ClpT:sol}) and (\ref{Clpn:sol}) 
into Eq. (\ref{Clp:ex}), we obtain
\begin{eqnarray}
C_\lambda^{\perp}(t_0,\tau) 
& = & -
\frac{\pi(d+2)^2(d-1)^2 T_0(t_0)^3 A}{ 2d^3 m\zeta^{*2}l_H^{d}}
\left (
\frac{ e^{-\zeta^* \tau}}{(2\pi(\eta^* + 2b)\tau)^{\frac{d+2}{2}}}
\right ),
\label{Clp:sol}
\end{eqnarray}
where $A$ is defined by Eq. (\ref{A:def}).

\subsection{The contribution from the longitudinal mode}
Substituting Eqs. (\ref{uparaex}) and (\ref{Tex}) into Eq. 
(\ref{Clpa:def}), we obtain the expression of $C_\lambda^{\parallel}(t_0,\tau)$ as
\begin{eqnarray}
C_\lambda^{\parallel}(t_0,\tau) 
& = &  
\frac{(d+2)n_H}{ 4l_H^{d}}
\int d\bv{v}_0 f_0(t_0,v_0) v_{0x}
\left ( mv_0^2 - (d+2)T_0(t_0) \right )
\int \frac{d\bv{k}}{(2\pi)^d} \hat{k}_x
 e^{-\left \{ \zeta + 2b k^2 \right \} \tau}
 \nonumber \\
& & \times 
 \left (
 -\frac{  i k \sqrt{T_0(t_0)} n_k(t_0)} {\zeta^{*}n_H\sqrt{m}}   
 +u_{k \parallel}(t_0)
 + \frac{i k (d-1)\sqrt{m} T_k(t_0)}{d\zeta^{*} \sqrt{m T_0(t_0)}}
 \right )
\nonumber \\
& & \times 
 \left (
 \frac{  (d-1) T_0(t_0) n_k(t_0) k^2} {dn_H\zeta^{*2}}   
 + \frac{ik (d-1) \sqrt{mT_0(t_0)} u_{k \parallel}(t_0)}{d\zeta^*}
 - \frac{(d-1)^2 T_k(t_0) k^2}{d^2\zeta^{*2}}
 \right ) \nonumber \\
 & = &
C_\lambda^{\parallel T}(t_0,\tau) + C_\lambda^{\parallel n}(t_0,\tau), \label{Clpa:ex}
\end{eqnarray}
where we introduce
\begin{eqnarray}
C_\lambda^{\parallel T}(t_0,\tau) 
& = &  -
\frac{(d+2)(d-1)^2n_H}{ 2d^2 \zeta^{*2}l_H^{d}}
\int d\bv{v}_0 f_0(t_0,v_0) v_{0x}
\left ( mv_0^2 - (d+2)T_0(t_0) \right ) 
\nonumber \\
& & \times
\int \frac{d\bv{k}}{(2\pi)^d} \hat{k}_x
u_{k \parallel x}(t_0)T_{-k}(t_0) k^2
 e^{-\left \{ \zeta + 2b k^2 \right \} \tau},
\label{ClpaT:def} \\
C_\lambda^{\perp n}(t_0,\tau) 
& = &  
\frac{(d+2)(d-1)T_0(t_0)}{ 2d \zeta^{*2}l_H^{d}}
\int d\bv{v}_0 f_0(t_0,v_0) v_{0x}
\left ( mv_0^2 - (d+2)T_0(t_0) \right )
\nonumber \\
& & \times
\int \frac{d\bv{k}}{(2\pi)^d} \hat{k}_x
u_{k \parallel }(t_0) n_{-k}(t_0) k^2
 e^{-\left \{ \zeta + 2b k^2 \right \} \tau}.
\label{Clpan:def}
\end{eqnarray}

Substituting Eq. (\ref{Initial:eq}) into Eq. (\ref{ClpaT:def}),
we obtain the expression for $C_{\lambda}^{\parallel T}(t_0,\tau)$
\begin{eqnarray}
C_\lambda^{\parallel T}(t_0,\tau) 
& = &  -
\frac{(d+2)(d-1)^2}{ 2d^3 n_H \zeta^{*2}l_H^{d}}
\int d\bv{v}_0 f_0(t_0,v_0) v_{0x}
\left ( mv_0^2 - (d+2)T_0(t_0) \right ) 
\nonumber \\
& & \times
\int \frac{d\bv{k}}{(2\pi)^d} 
\hat{k}_x (\bv{v}_0 \cdot \hat{\bv{k}})
(mv_0^2 - dT_0(t_0)) |W_k|^2 k^2
 e^{-\left \{ \zeta + \left ( \frac{1}{2} \eta^* +b \right )k^2 \right \} \tau}
 \nonumber \\
& \simeq &  -
\frac{(d+2)^2(d-1)^2 T_0(t_0)^3}{ 2d^3 m\zeta^{*2}l_H^{d}}
\int \frac{d\bv{k}}{(2\pi)^d} 
 \hat{k}_x^2 k^2
 e^{-\left \{ \zeta + 2b k^2 \right \} \tau}
 \nonumber \\
& = &  -
\frac{\pi(d+2)^2(d-1)^2 T_0(t_0)^3}{ d^3 m\zeta^{*2}l_H^{d}}(2+a_2^H(d+10))
\left (
\frac{ e^{-\zeta^* \tau}}{(8\pi b \tau)^{\frac{d+2}{2}}}
\right ).
\label{ClpaT:sol}
\end{eqnarray}
For $C_{\lambda}^{\parallel n}(t_0,\tau)$, by substituting Eq. (\ref{Initial:eq}) into Eq. (\ref{Clpan:def}),
we obtain 
\begin{eqnarray}
C_\lambda^{\parallel n}(t_0,\tau) 
& = & 
\frac{(d+2)(d-1)T_0(t_0)}{ 2d n_H \zeta^{*2}l_H^{d}}
\int d\bv{v}_0 f_0(t_0,v_0) v_{0x}
\left ( mv_0^2 - (d+2)T_0(t_0) \right ) 
\nonumber \\
& & \times
\int \frac{d\bv{k}}{(2\pi)^d} 
\hat{k}_x \bv{v}_0 \cdot \hat{\bv{k}}
|W_k|^2 k^2
 e^{-\left \{ \zeta + 2b k^2 \right \} \tau}
 \nonumber \\
& \simeq & 
\frac{(d+2)^2(d-1)T_0(t_0)^3 a_2^H}{ 2d m\zeta^{*2}l_H^{d}}
\int \frac{d\bv{k}}{(2\pi)^d} 
\hat{k}_x^2 k^2
 e^{-\left \{ \zeta + 2b k^2 \right \} \tau}
 \nonumber \\
& = & 
\frac{\pi(d+2)^2(d-1)^2 T_0(t_0)^3 a_2^H}{ d m\zeta^{*2}l_H^{d}}
\left (
\frac{ e^{-\zeta^* \tau}}{(8\pi b \tau)^{\frac{d+2}{2}}}
\right ).
\label{Clpan:sol}
\end{eqnarray}
Thus, from Eqs. (\ref{Clpa:ex}), (\ref{ClpaT:sol}) and (\ref{Clpan:sol}) 
 we obtain $C_{\lambda}^{\parallel}(t_0,\tau)$ as
\begin{eqnarray}
C_\lambda^{\parallel}(t_0,\tau) 
& = & -
\frac{\pi(d+2)^2(d-1)^2 T_0(t_0)^3 A}{ 2d^3 m\zeta^{*2}l_H^{d}}
\left (
\frac{ e^{-\zeta^* \tau}}{(8\pi b \tau)^{\frac{d+2}{2}}}
\right ).
\label{Clpa:sol}
\end{eqnarray}
Finally, from Eqs. (\ref{Clp:sol}) and (\ref{Clpa:sol})
associated with Eq. (\ref{Cl:ex}), we obtain Eq. (\ref{Cl:sol}).

\end{document}